# Second harmonic generation by strongly coupled exciton-plasmons: the role of polaritonic states in nonlinear dynamics

Maxim Sukharev[1,2,*], Adi Salomon[3], Joseph Zyss[4]

[1]College of Integrative Sciences and Arts, Arizona State University, Mesa, AZ 85212, USA

[2]Department of Physics, Arizona State University, Tempe, AZ, USA

[3]Department of Chemistry, Institute of Nanotechnology and Advanced Materials (BINA), Bar-Ilan University, Ramat-Gan 5290002, Israel

[4] LUMIN Laboratory and Institut d'Alembert, Ecole Normale Supérieure Paris Saclay, CNRS, Université Paris-Saclay, 4, avenue des Sciences, Gif-sur-Yvette, France

*corresponding author: maxim.sukharev@asu.edu

We investigate second harmonic generation (SHG) from hexagonal periodic arrays of triangular nano-holes of aluminum using a self-consistent methodology based on the hydrodynamics-Maxwell-Bloch approach. It is shown that angular polarization patterns of the far-field second harmonic response abide to three-fold symmetry constraints on tensors. When a molecular layer is added to the system and its parameters are adjusted to achieve the strong coupling regime between a localized plasmon mode and molecular excitons, Rabi splitting is observed from occurrence of both single- and two-photon transition peaks within the SHG power spectrum. It is argued that the splitting observed for both transitions results from direct transitions between lower and upper polaritonic states of the strongly coupled system. This interpretation can be accounted by a tailored three-level quantum model, with results in agreement with the unbiased numerical approach. Our results suggest the hybrid states formed in strongly coupled systems directly contribute to the nonlinear dynamics. This opens new directions in designing THz sources and nonlinear frequency converters.

**Introduction**

The research in plasmonics found various applications in chemistry,[1,2] biology,[3] engineering.[4,5] All are due to the ability of plasmonic materials to sustain plasmon-polariton resonances with a very small mode volume.[6] Even though characteristic Q-factors of such modes are relatively small large local field enhancement[7] makes such systems very attractive for numerous applications including nonlinear nano-optics[8,9] and plasmon-polariton chemistry.[10] The latter combines small plasmon mode volumes with molecular excitons.[11] When the coupling strength of a local field and an exciton surpasses all the damping rates in a system the strong coupling regime is reached. In such a regime new hybrid states called upper and lower polaritons are formed.[12] Under strong coupling conditions these states have a mixed plasmon-exciton character.

While the linear optics of such systems is being actively explored the nonlinear phenomena such as second harmonic generation (SHG) recently began attracting considerable attention.[13-17] What makes nonlinear plasmonics an exciting playground in terms of fundamental as well as applied objectives? From a fundamental viewpoint, interactions of organic and inorganic systems, including molecules of various complexity with metals, challenges our abilities to properly describe such systems using conventional semi-classical methodology. At the application side, advanced sensing and bio-sensing devices relying on the sensitivity of nonlinear effects to perturbations and on the unlimited possibilities for targeted surface sensitization open-up new horizons in diagnosis and therapy.

In this manuscript we report theoretical investigations of the nonlinear properties of a strongly coupled exciton-plasmon system. It is comprised of a planar array of metallic nanoscale cavities with a contacting molecular layer of a general nature. Whereas SHG has been chosen as the central target at this stage, other phenomena can be predicted along the same theoretical lines, such as other three- and four-wave mixing schemes, among which third-harmonic generation, intensity dependent index of refraction, parametric effects including near IR and THz generation.

Our approach is based on the numerical exploitation of a non-perturbative hydrodynamic-Maxwell model. It can be therefore viewed to some extent as a "numerical experiment", then calling, very much as in the case of an actual experiment, for simple physical interpretations of numerical findings. Interpretations of the numerical data are based on a semi-classical quantum picture relying on a minimal three-level system that appears sufficient to account for the main features obtained from simulations. The proposed analytical model allows for simple expressions that are prone to straightforward interpretations in terms of photon mediated coupling between the basis states. The purpose of our studies is to theoretically explore in a relatively generic way the potential for nonlinear optics of mixed metal-molecule structures where strong interactions between both species are expected to display new features with the potential to enhance their efficiency based on well identified parameters.

Early theoretical works in nonlinear plasmonics combined either a nonlinear material with otherwise linear plasmonic interface,[18] employed ab initio calculations limited due to obvious reasons to small sizes[19], or utilized a semi-classical hydrodynamic model for conduction electrons.[20] The latter has recently proved to be highly effective even on a quantitative level.[21] It was recently demonstrated[15] that combining the hydrodynamic model for metal with Bloch equations for molecules in time domain one can truly explore nonlinear optics of strongly coupled systems. The major question that we attempt to address in this work is: in a strongly coupled system with two well-defined polaritonic states, can we expect to observe the hybrid states directly participating in nonlinear dynamics? And if so, are there direct lower-to-upper polariton transitions? The positive answer to this question would lead to a large body of exciting applications ranging from THz generation to electromagnetic induced transparency and slow light.[22]

In the first part of our manuscript, we overview the main features of the hydrodynamic-Maxwell-Bloch model. Next, we discuss our main results based on this model in terms of linear and nonlinear spectroscopic as well as polarization properties. Finally, we propose a simple analytical three-level quantum model and confront it to the former "numerical experiment" and then draw conclusions. Additionally, we provide appendix, which discusses polarization properties of SHG, based on abstract tensor and symmetry considerations as well as a more visual geometric argumentation.

**Theory**

We consider periodic arrays of triangular nanoholes arranged in a hexagonal lattice as schematically depicted in Fig. 1a. Optical response of this system is evaluated using Maxwell's equations in the space-time domain

$$\dot{\mathbf{B}} = -\nabla \times \mathbf{E},$$
$$\varepsilon_0 \dot{\mathbf{E}} = \frac{1}{\mu_0} \nabla \times \mathbf{B} - \dot{\mathbf{P}}, \quad (1)$$

where the macroscopic polarization **P** is determined by the hydrodynamic model of conduction electrons[23]. The equation of motion describing **P** combines both the continuity equation relating the polarization current and the number density, $n$, and the equation determining dynamics of the electron velocity field.[24] Such an equation can be written in a concise form[25,26]

$$\ddot{\mathbf{P}} + \gamma_e \dot{\mathbf{P}} = \frac{n_0 e^2}{m_e^*} \mathbf{E} + \frac{e}{m_e^*}\left(\dot{\mathbf{P}} \times \mathbf{B} - \mathbf{E}(\nabla \cdot \mathbf{P})\right) - \frac{1}{n_0 e}\left((\nabla \cdot \dot{\mathbf{P}})\dot{\mathbf{P}} + (\dot{\mathbf{P}} \cdot \nabla)\dot{\mathbf{P}}\right), \qquad (2)$$

here $m_e^*$ is the effective mass of a conduction electron, $n_0$ is the equilibrium number density of electrons, and $\gamma_e$ is the phenomenological damping rate. In the simulations we parametrize the plasma frequency, $\Omega_p = \sqrt{n_0 e^2 / \varepsilon_0 m_e^*}$, the number density, and the damping rate. In this paper we consider aluminum with the following parameters: $\Omega_p = 10.85$ eV, $\gamma_e = 0.4641$ eV, and $n_0 = 1.81 \times 10^{29}$ m$^{-3}$, the geometrical parameters defining the array are shown in the caption of Fig. 1.

In addition to the metal, we also consider two-level molecules forming a thin layer on the input side of the array. The molecular layer is simulated using rate equations[12] that determine the dynamics of the number density of molecules in the excited state, $N_2$, and the corresponding macroscopic polarization of the molecular layer, **P**,

$$\begin{aligned}
\dot{N}_2 + \gamma_{21} N_2 &= \frac{1}{\hbar \omega_{12}} \mathbf{E} \cdot \dot{\mathbf{P}}, \\
\ddot{\mathbf{P}} + \Gamma \dot{\mathbf{P}} + \omega_{12}^2 \mathbf{P} &= \frac{2}{3\hbar} \omega_{12} \mu_{12}^2 (N_0 - 2N_2) \mathbf{E},
\end{aligned} \qquad (3)$$

where $\gamma_{21}$ is the population relaxation rate, $\Gamma = \gamma_{21} + 2\gamma_d$, $\gamma_d$ is the pure dephasing rate, $\omega_{12}$ and $\mu_{12}$ are the transition frequency and the transition dipole, respectively, and $N_0$ is the total number density of molecules. The initial conditions correspond to all molecules in the ground state. The parameters used in this work are: $\gamma_{21} = 4.14$ meV, $\gamma_d = 7.03$ meV, $\mu_{12} = 10$ Debye, $N_0 = 2 \times 10^{26}$ m$^{-3}$, the thickness of the layer is 20 nm.

Eqs. (1) – (3) are discretized in space and time following finite-difference time-domain method.[15] When evaluating both linear and nonlinear responses the system is excited by a planewave at normal incidence. We follow the conventional total field/scattered field approach[27] to generate plane waves. Periodic boundary conditions are applied along X and Y (see Fig. 1a) and convolution perfectly matched layers[28] are set on top and bottom of the grid to absorb outgoing waves. The EM energy flux is detected on the input (reflection) and output (transmission) sides of the array in the far-field zone as the corresponding Poynting vector integrated over a given detection plane. In order to speed the simulations up we employ a three-dimensional domain decomposition technique and message-passing interface subroutines. Our home-build codes are parallelized for 12×12×8 = 1152 processors with a spatial resolution of 1.5 nm and a time step of 2.5 attoseconds. Linear spectra shown in Figs. 1 and 3 are usually converged after propagating equations for 300 fs. The average executions times vary between 15 to 30 minutes. The nonlinear simulations are performed using a 500 fs long pump. To eliminate spurious nonlinear contributions that always occur due to spatial discretization we apply Blackman-Harris time window to all detected fields.[19] Such a procedure results in clear power spectra with well-resolved harmonics. The execution times of our nonlinear codes are between 2 and 3 and a half hours.

**Results and discussion**

Figs. 1b and 1c present linear transmission and reflection spectra of the nanohole array in vacuum for different thicknesses of the metal film. We note that both transmission and reflection exhibit multiple resonances with the broad mode clearly demonstrating the splitting, which is more pronounced in the reflection. The splitting is gradually disappearing with increasing thickness indicating that the physics of this is due to the coupling between surfaces modes on the input and output sides of the array forming symmetric and antisymmetric states with respect to the center of the film (such modes were first discussed in Ref. [29] and then later in Ref. [30], experimental implications of these modes and their properties were examined in Ref. [31]). The broad mode with the splitting centered at 2.45 eV corresponds to a localized plasmon mode with EM field primarily localized near the vertices of holes. The other modes observed near 3 eV are Bragg-plasmon modes. In this paper we consider localized plasmon mode, its coupling to molecules, and is influence on SHG. We note that the localized plasmon enhances SHG efficiency noticeably higher compared to Bragg plasmons as demonstrated in Refs. [15, 17].

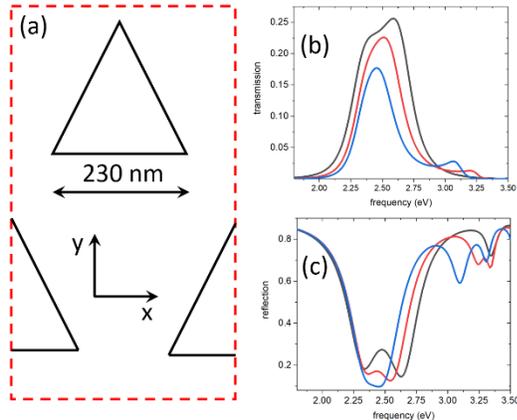

**Fig. 1.** Linear spectra of the periodic hexagonal array of triangular nanoholes in Al metal film. Panel (a) shows schematic setup of the array. Holes have a shape of equilateral triangles with a side of 230 nm. Red dashed lines indicate periodic boundaries. Panels (b) and (c) show linear transmission and reflection as functions of the incident frequency, respectively. Black lines show results for the 250 nm thick metal. Red lines are for 300 nm thickness, and blue lines are for 350 nm thick film. The period of the array is 395 nm.

Fig. 2 shows angular diagrams of SHG response detected in the transmitted signal. We perform two types of detection: (1) calculating the SHG signal polarized along the pump polarization and perpendicular to it (Figs. 2a and 2b); (2) calculating the SHG signal polarized along X and Y axis in the fixed (laboratory) system of coordinates (see Fig. 1a where X and Y are defined).

Fig. 2 displays two strikingly different SHG polarization patterns. In panels (a) and (b), six lobes in sixfold symmetry configuration whereas panels (c) and (d) exhibit four lobes in a four-fold symmetry configuration. These angular symmetry features are discussed in great details in Appendix based on the tensor symmetry considerations along with a pictorial discussion of the non-intuitive fourfold emission from a threefold symmetric nonlinear response.

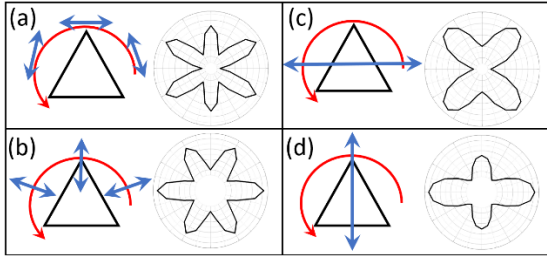

**Fig. 2.** Angular properties of the second harmonic generation enhanced by a localized plasmon mode. Angular diagrams correspond SHG signal detected on the output side of the array as a function of the pump polarization. Panels (a) and (b) correspond to the detector that rotates together with the pump. SHG polarized parallel to the pump is shown in panel (a). Panel (b) shows SHG polarized perpendicular to the pump. Panels (c) and (d) correspond to the stationary detector in the laboratory system of coordinates. Horizontally polarized SHG is shown in panel (c) and the vertically polarized SHG is in panel (d). Simulations are performed for the array with the period of 395 nm. Metal thickness is 350 nm. The pump frequency is 2.45 eV.

It is informative to examine the SHG signal as a function of the pump frequency. In general, any dispersive material with a second order nonlinearity exhibits the second-order susceptibility, which can be written as a product of the first-order susceptibility at the fundamental and second harmonic frequencies following Miller's rule[32]

$$\chi^{(2)}(2\omega;\omega,\omega) \sim \chi^{(1)}(2\omega)\left(\chi^{(1)}(\omega)\right)^2. \tag{4}$$

We thus should expect to observe peaks in the SHG signal corresponding to one- and two-photon emission. Fig. 3 presents the conversion efficiency in transmission and reflection as a function of the pump frequency. The conversion efficiency defined as the ratio of the intensity of the second harmonic signal, $I(2\omega)$, and the intensity at the fundamental frequency, $I(\omega)$.

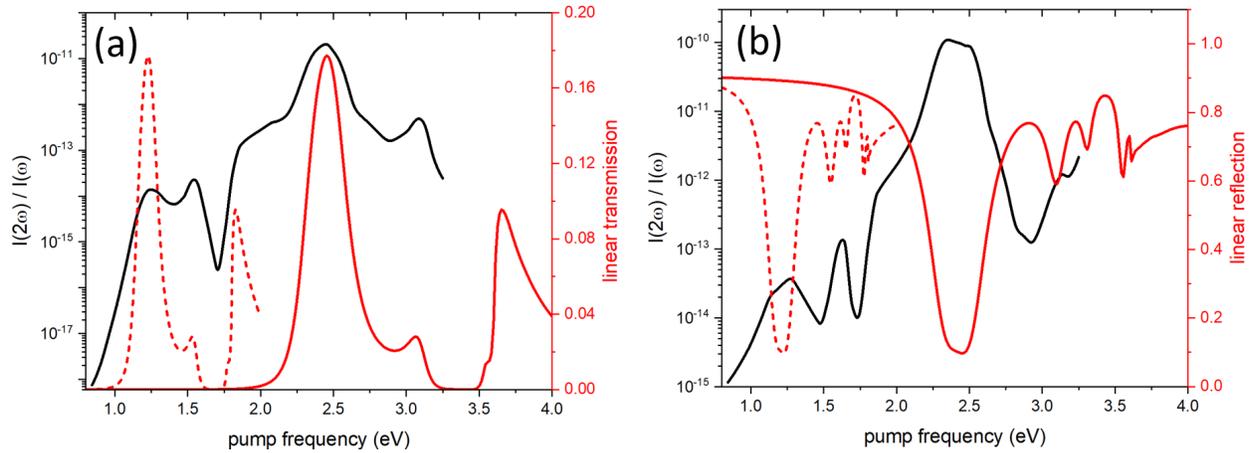

**Fig. 3.** Conversion efficiency defined as a ratio of the intensity of the second harmonic and the intensity at the fundamental frequency as a function of the pump frequency. Panel (a) shows data for the transmission. Panel (b) shows the reflection. Both panels present data on a double-Y scale. SHG efficiency is shown as black (in a logarithmic scale) and the linear transmission/reflection are shown as red lines. The red dashed lines correspond to transmission/reflection plotted as a function of the half-frequency. Material parameters are the same as in Fig. 2.

The results presented in Fig. 3 demonstrate Miller's rule for plasmonic systems. There are clear signatures of both one- and two-photon emission enhanced by corresponding plasmon modes of the array. As expected, the largest enhancement of the conversion efficiency is due to the localized plasmon at 2.45 eV, which is higher than that due to the Bragg plasmon at 3.1 eV by more than an order of magnitude as was reported elsewhere.[15] The two-photon emission peaks also have an obvious correspondence to the plasmon modes of the array when compared with the dashed red line. Interestingly Wood's anomaly, i.e., the peak associated with the period of the array,[33] influences the conversion efficiency as seen near 2 eV as a knee shape of the conversion efficiency. Another observation worth emphasizing is the splitting of the plasmon peak discussed in Figs. 1b and 1c, which is mostly pronounced in the reflection near 2.4 eV. One can note that the two-photon emission peak at 1.2 eV also evidences a clear signature of the splitting. Interestingly, the splitting being due to the hybridization of the surface modes on the input and output sides of the array has a different structure in the one-photon and two-photon peaks. Notice the low frequency mode has a higher conversion efficiency at the one-photon peak compared to the higher frequency mode. The two-photon peak has this reversed. This again reinforces the hypothesis that the local symmetry plays an important role in the SHG process.

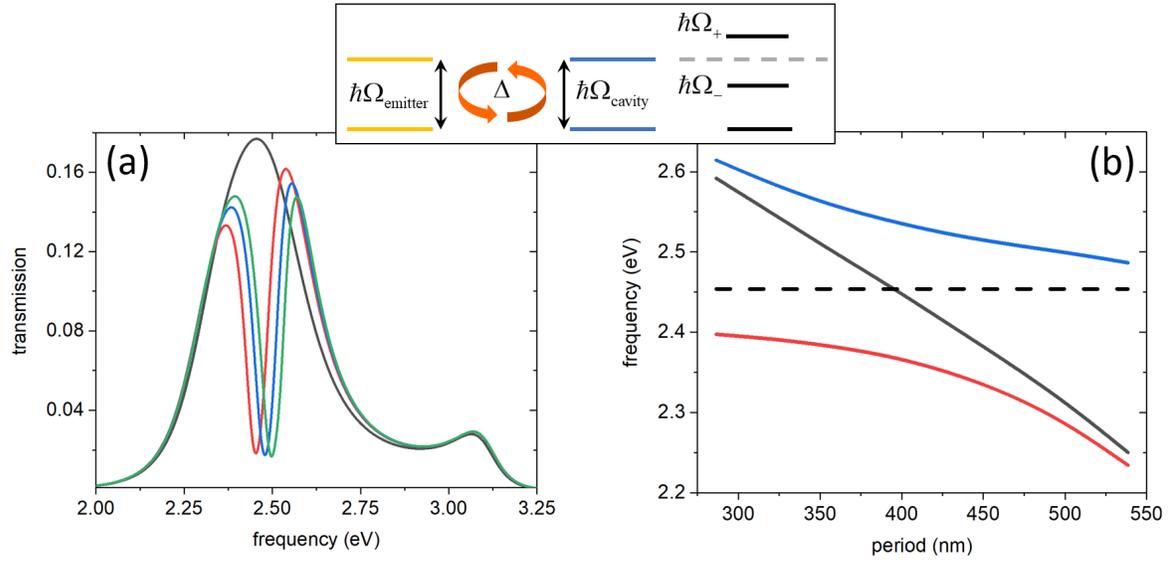

**Fig. 4.** The inset shows a schematic energy diagram describing the strong coupling between an emitter and an optical cavity with a formation of two polaritonic bands. The main panel shows the linear optical response of a 20 nm thin molecular layer on top of the periodic array of triangular nano-holes. Panel (a) shows transmission as a function of the incident frequency. Black line shows transmission for the array with no molecules; red line shows data for molecules with $\omega_{12}$ = 2.45 eV; blue line is for $\omega_{12}$ = 2.48 eV; and green line is for $\omega_{12}$ = 2.50 eV. Panel (b) shows frequencies of the upper (blue) and lower (red) polaritons as functions of the period of the array. Black line shows the frequency of the localized plasmon mode and the black dashed line indicates the transition frequency of the molecules (fixed at 2.45 eV).

We now add a thin molecular layer placing it on top of the array and examine its influence on the SHG. Fig. 4 explores the linear optical response of such a system. The transmission with molecules being resonant with the localized plasmon mode exhibits a clear Rabi splitting indicative of a strong coupling between molecular excitons and the corresponding plasmon mode. Fig. 4b shows frequencies of the upper and lower polaritons as functions of the periodicity of the array exhibiting expected avoided crossing. The Rabi splitting observed for the array with a period of 395 nm is measured to be 179 meV. Next, we explore the SHG process in such a system under strong coupling conditions.

The main result of this manuscript is shown in Fig. 5. Here we plot the SHG efficiency for the molecules/metal system under strong coupling conditions and compare it directly with the results obtained without molecules. Unexpectedly the two-photon peak exhibits a clear signature of the Rabi splitting. Previously it was argued that the Rabi splitting observed in the second harmonic signal for strongly coupled systems can be explained by expanding the two coupled oscillators model.[14,16] *With molecules treated as simple two-level emitters with no diagonal elements of the transition dipole the only source of the second harmonic is the metal.* The model based on a strongly driven oscillator with a second order nonlinearity (plasmons) coupled to a linear oscillator (molecular excitons) results in the modified second-order susceptibility (5).[14] Only the second term in (5) carries the signature of the strong coupling

$$\chi^{(2)} \sim \left( \frac{1}{\omega_{SPP}^2 + i\gamma_{SPP}(2\omega) - (2\omega)^2} \right) \cdot \left( \frac{\omega_{exc}^2 + i\gamma_{exc}\omega - \omega^2}{g^4 - \left(\omega_{SPP}^2 + i\gamma_{SPP}\omega - \omega^2\right)\left(\omega_{exc}^2 + i\gamma_{exc}\omega - \omega^2\right)} \right)^2, \quad (5)$$

where $g$ is the coupling strength between the plasmon (SPP) and the corresponding molecular exciton (exc). Eq. (5) evidently predicts that the Rabi splitting does not appear in the two-photon peak since only plasmon enhancement is observed.

In contrast with the coupled oscillators model our results shown in Fig. 5a indicate that the

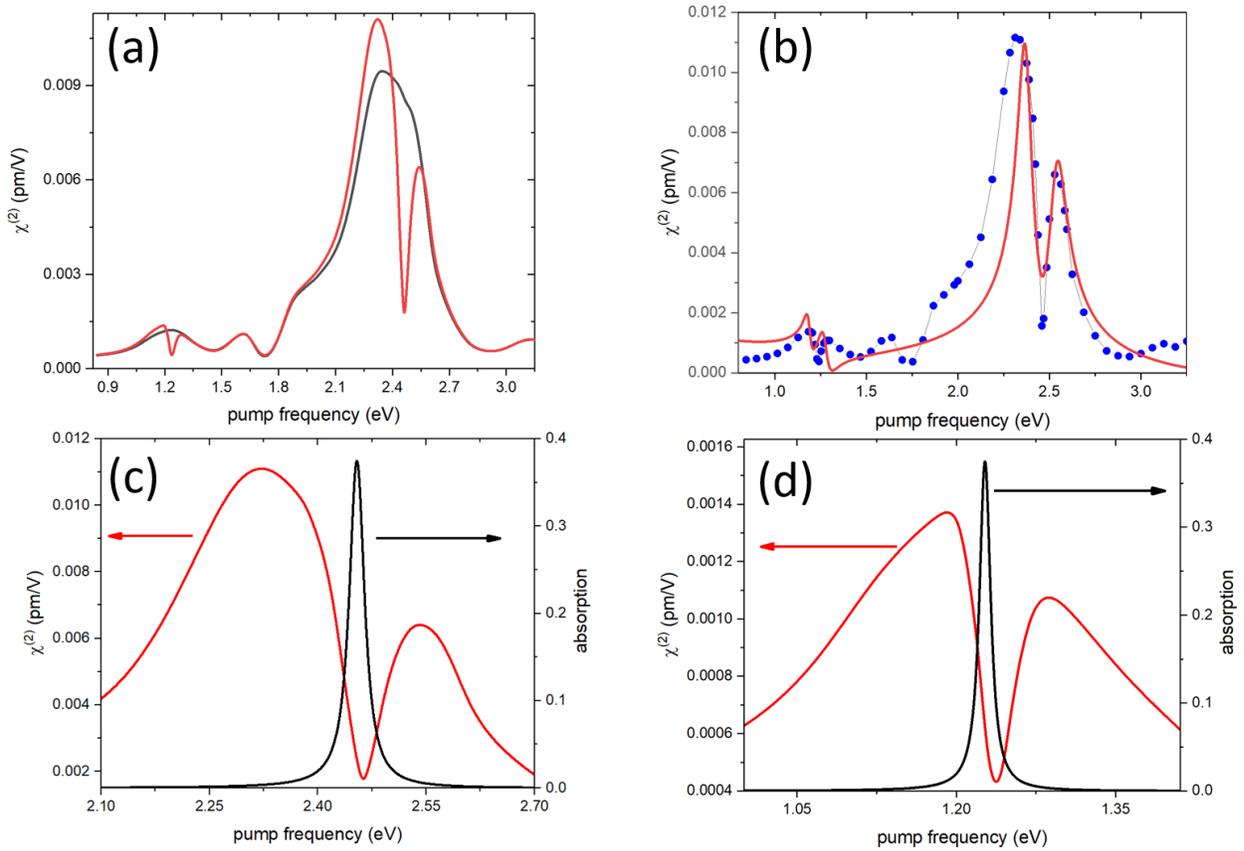

**Fig. 5.** Effective second-order susceptibility $\chi^{(2)}$ as a function of the pump frequency. Panel (a) shows $\chi^{(2)}$ for metal only (black line) and metal with molecules (red line). Panel (b) shows $\chi^{(2)}$ (blue circles) and the fitting using expression (6) (red line). Panels (c) and (d) show $\chi^{(2)}$ (red line) and linear absorption due to molecular layer only (black line). Panel (c) is a zoom-in of the frequency region near one-photon emission and panel (d) corresponds to the two-photon peak (linear absorption here plotted as a function of the half-frequency). The molecular transition frequency is 2.45 eV and the periodicity is 395 nm.

molecules also contribute to the two-photon emission. One may argue that the splitting seen near

1.2 eV may correspond to the absorption of two 1.2 eV photons by the molecules. However, the splitting observed at 1.2 eV equals exactly the half of the Rabi splitting, while the absorption linewidth for molecules is significantly smaller. This is also illustrated in Figs. 5c and 5d, where we plot linear absorption of the molecular layer right next to the corresponding one- and two-photon peaks. One can see that the linear absorption is significantly narrower and noticeably red-shifted compared to the observed resonances. Another interpretation of the observed splitting at 1.2 eV is thus needed to account for the $2\omega$ splitting.

We propose a three-level quantum model in the semi-classical approximation to describe a two band plasmon-polariton spectrum close to the avoided crossing point. 0 designates the ground state, 1 denotes the lower polaritonic branch with energy $\hbar\omega_1$, and 2 denotes the upper one with energy $\hbar\omega_2$. We limit ourselves to the component $\chi^{(2)}_{yyy}$ for threefold symmetry as the only other component $\chi^{(2)}_{yxx}$ can be obtained via $\chi^{(2)}_{yyy} + \chi^{(2)}_{yxx} = 0$. We also assume that the transitions are only weakly populating the excited states, making the population dominantly in the ground state. In the density matrix formulation,[32] the only remaining diagonal term is $\rho^{(0)}_{00} = 1$ thus neglecting excited state populations. In the following, we shall omit the cartesian indices which will all be implicitly taken as *yyy*.

We can now proceed to adopt the general expressions given (page 174, Eq. (3.6.16) in Ref. [32]) in the case of a non-dipolar three-level system. For a *N*-level system we have

$$\chi^{(2)}(-2\omega;\omega,\omega) = \frac{N}{2\hbar^2} \sum_{n,m} (a+b),$$

$$a = \frac{\mu_{0n}\mu_{nm}\mu_{m0}}{(\omega_n - 2\omega - i\gamma_n)(\omega_m - \omega - i\gamma_m)}, \quad (6)$$

$$b = \frac{\mu_{0n}\mu_{nm}\mu_{m0}}{(\omega_{nm} + 2\omega + i\gamma_n)(\omega_m - \omega - i\gamma_m)},$$

where each term in *a* and *b* corresponds to a three-photon transition from the ground state and back, with any pair of intermediate excited states chosen among the *N* states.

Reduction to a three-level system limits the summation to *n* = 1, 2, likewise for *m*. Moreover, threefold symmetry and the absence of dipolar properties in the molecular layer leads to the cancellation of all participating dipoles in the summation, resulting for the case of a three-level system in $\mu_{00} = \mu_{11} = \mu_{22} = 0$. The *a* and *b* terms take then the following simplified expressions

$$a = \frac{\mu_{01}\mu_{12}\mu_{20}}{(\omega_1 - 2\omega - i\gamma_1)(\omega_2 - \omega - i\gamma_2)} + \frac{\mu_{02}\mu_{21}\mu_{10}}{(\omega_2 - 2\omega - i\gamma_2)(\omega_1 - \omega - i\gamma_1)},$$

$$b = \frac{\mu_{01}\mu_{12}\mu_{20}}{(\omega_{12} + 2\omega + i\gamma_{12})(\omega_2 - \omega - i\gamma_2)} + \frac{\mu_{02}\mu_{21}\mu_{10}}{(\omega_{21} + 2\omega + i\gamma_{21})(\omega_1 - \omega - i\gamma_1)},$$ (7)

where $\omega_{12} = \omega_1 - \omega_2 < 0$ and $\omega_{21} = \omega_2 - \omega_1 > 0$.

The two denominators of the term clearly exhibit two and one-photon resonances respectively at $\omega_1$ and $\omega_2$ in agreement with the *a priori* non-perturbative fully vectorial hydrodynamic-Maxwell-Bloch model. This behavior corresponds to one- and two-photon resonant second harmonic generation processes for a pair of two-level systems, namely $(|0\rangle, |1\rangle)$ and $(|0\rangle, |2\rangle)$. Using simplified Eq. (7) we performed fitting of the second-order susceptibility to numerical results. The result is shown in Fig. 5b. Although we note that the fitting is obviously not perfect the model adequately describes splitting observed at one- and two-photon peaks.

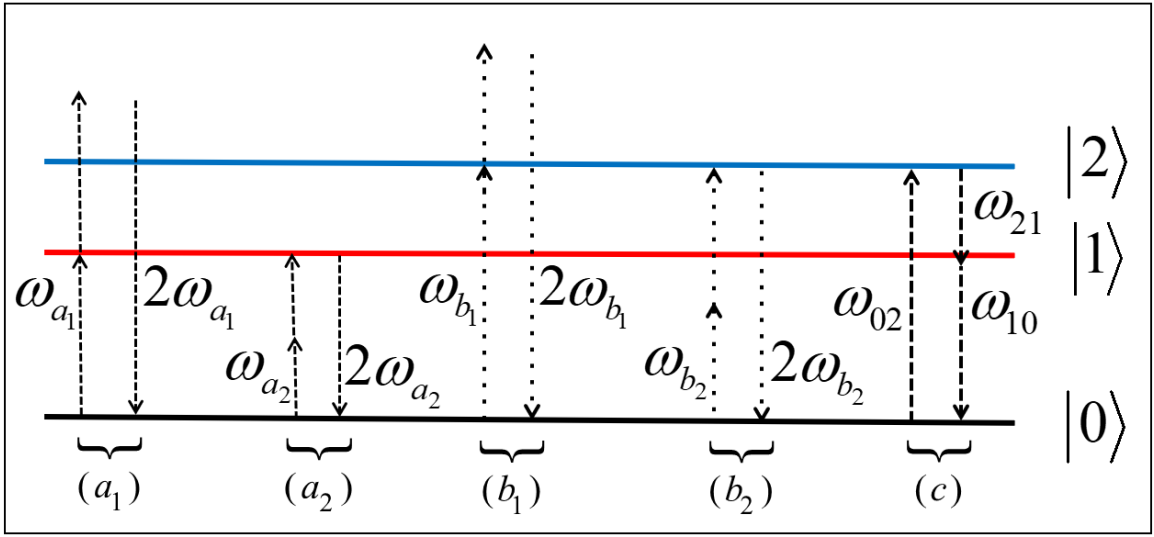

**Fig. 6.** Transition schemes for SHG and DFG processes within a three-level quantum system as discussed in the text. Schemes $(a_1)$ and $(b_1)$ correspond to one photon resonances in SHG whereas $(a_2)$ and $(b_2)$ depict two-photon resonances. Scheme (c) describes a difference frequency mixing process, which sustains parametric emission or difference frequency generation depending on the experimental configuration (namely only one incoming beam at frequency $\omega_{02}$ for parametric processes and two beams respectively at $\omega_{01}$ and $\omega_{02}$ for the difference mixing process, which can also be considered as a stimulated parametric processes). Scheme (c) is seen to provide the benefit of a doubly resonant process.

The four possible corresponding schemes are sketched in Fig. 6 under $(a_{1,2})$ and $(b_{1,2})$. Following conventions on Fig. 6, $\omega_{a_1}$ and $\omega_{b_1}$ are in one-photon resonance with levels $|1\rangle$ and

$|2\rangle$ whereas $2\omega_{a_2}$ and $2\omega_{b_2}$ are in two-photon resonance with the same levels. In contrast with expression $a$ the two denominators in $b$ exhibit only one-photon resonances respectively at both $\omega_1$ and $\omega_2$.

The b term does not exhibit resonance in the optical regime since $\omega_{21} \ll \omega, 2\omega$. However, a modified b term could be of interest in the context of a possible difference-frequency generation (*DFG*) process $\omega' \to (\omega, \Omega = \omega' - \omega)$, inverse to the corresponding $(\omega, \Omega) \to \omega + \Omega$ sum-frequency generation (*SFG*) process, whereby a coherent beam at frequency $\Omega$ in the THz regime could be generated in resonance with the $\omega_{21}$ transition. The corresponding susceptibility can be evaluated in two steps: the first one by way of generalizing the *SHG* susceptibility as detailed below onto *SFG* $(\omega_p, \omega_q) \to \omega_p + \omega_q$ then replacing $\omega_p$ by $-\omega_p$ corresponding to *DFG*.

For *SFG* we have

$$\chi^{(2)}(-\omega_p - \omega_q; \omega_p, \omega_q) = \frac{N}{2\hbar^2}(a_1 + a_2 + b_1 + b_2),$$

$$a_1 = \frac{\mu_{01}\mu_{12}\mu_{20}}{(\omega_2 - \omega_p - \omega_q - i\gamma_2)(\omega_1 - \omega_p - i\gamma_1)} + \frac{\mu_{02}\mu_{21}\mu_{10}}{(\omega_1 - \omega_p - \omega_q - i\gamma_1)(\omega_2 - \omega_p - i\gamma_2)},$$

$$a_2 = \frac{\mu_{01}\mu_{12}\mu_{20}}{(\omega_2 - \omega_p - \omega_q - i\gamma_2)(\omega_1 - \omega_q - i\gamma_1)} + \frac{\mu_{02}\mu_{21}\mu_{10}}{(\omega_1 - \omega_p - \omega_q - i\gamma_1)(\omega_2 - \omega_q - i\gamma_2)}, \quad (8)$$

$$b_1 = \frac{\mu_{01}\mu_{12}\mu_{20}}{(\omega_{21} + \omega_p + \omega_q + i\gamma_{21})(\omega_1 - \omega_p - i\gamma_1)} + \frac{\mu_{02}\mu_{21}\mu_{10}}{(-\omega_{21} + \omega_p + \omega_q + i\gamma_{21})(\omega_2 - \omega_p - i\gamma_2)},$$

$$b_2 = \frac{\mu_{01}\mu_{12}\mu_{20}}{(\omega_{21} + \omega_p + \omega_q + i\gamma_{21})(\omega_1 - \omega_q - i\gamma_1)} + \frac{\mu_{02}\mu_{21}\mu_{10}}{(-\omega_{21} + \omega_p + \omega_q + i\gamma_{21})(\omega_2 - \omega_q - i\gamma_2)}.$$

For *DFG* we obtain

$$\chi^{(2)}(\omega_p - \omega_q; -\omega_p, \omega_q) = \frac{N}{2\hbar^2}(\alpha_1 + \alpha_2 + \beta_1 + \beta_2),$$

$$\alpha_1 = \frac{\mu_{01}\mu_{12}\mu_{20}}{(\omega_2 + \omega_p - \omega_q - i\gamma_2)(\omega_1 + \omega_p - i\gamma_1)} + \frac{\mu_{02}\mu_{21}\mu_{10}}{(\omega_1 + \omega_p - \omega_q - i\gamma_1)(\omega_2 + \omega_p - i\gamma_2)},$$

$$\alpha_2 = \frac{\mu_{01}\mu_{12}\mu_{20}}{(\omega_2 + \omega_p - \omega_q - i\gamma_2)(\omega_1 - \omega_q - i\gamma_1)} + \frac{\mu_{02}\mu_{21}\mu_{10}}{(\omega_1 + \omega_p - \omega_q - i\gamma_1)(\omega_2 - \omega_q - i\gamma_2)}, \quad (9)$$

$$\beta_1 = \frac{\mu_{01}\mu_{12}\mu_{20}}{(\omega_{21} - \omega_p + \omega_q + i\gamma_{21})(\omega_1 + \omega_p - i\gamma_1)} + \frac{\mu_{02}\mu_{21}\mu_{10}}{(-\omega_{21} - \omega_p + \omega_q + i\gamma_{21})(\omega_2 + \omega_p - i\gamma_2)},$$

$$\beta_2 = \frac{\mu_{01}\mu_{12}\mu_{20}}{(\omega_{21} - \omega_p + \omega_q + i\gamma_{21})(\omega_1 - \omega_q - i\gamma_1)} + \frac{\mu_{02}\mu_{21}\mu_{10}}{(-\omega_{21} - \omega_p + \omega_q + i\gamma_{21})(\omega_2 - \omega_q - i\gamma_2)}.$$

We further chose the incoming frequencies close to each other, such that the frequency difference $\omega_q - \omega_p$ be much smaller than optical frequencies, namely $\omega_q - \omega_p \ll \omega_1, \omega_2$ and of the order of the frequency shift between the higher and lower polaritonic branches, i.e. $\omega_q - \omega_p \simeq \omega_{21}$ which can span the THz domain.

Under such conditions, the $\alpha_1$ expression cannot support any resonances at optical frequencies, as it would otherwise contradict the above assumptions. The $\alpha_2$ expression can be resonant respectively at $\omega_q = \omega_1$ for its first term or at $\omega_q = \omega_2$ for the second one.

More interesting contributions are arising from the $\beta_1$ and $\beta_2$ expressions, i.e. resonances at $\omega_q - \omega_p = \omega_{21}$. In addition, $\omega_q = \omega_2$ in the second term of $\beta_2$ can lead to a second resonance. Such a doubly resonant scheme is displayed under (c) in Fig. 6. Whereas such a scheme would be probably of limited interest in a bulk medium with the benefit of a double resonance offset by the cost of corresponding absorptions for each resonance, the situation appears much more favorable for surface driven nonlinearities at the nanoscale where propagative absorption is negligible at the scale of the structure.

We note that the doubly resonant process described above could be of major interest for the efficient generation as well as detection of THz radiation, the structure studied here playing the role of a nanoscale emitter-receptor antenna in the THz range.

At a more fundamental level, such a process would be also of major spectroscopic interest as it could permit to resonantly access the otherwise elusive $\mu_{12}$ transition dipole that couples levels 1 and 2. In the generic octupolar molecular system described in Ref. [34], levels $|1\rangle$ and $|2\rangle$ then referred to as $|\bar{1}\rangle$ are degenerate for essential symmetry reasons resulting from threefold symmetry. Our present plasmonic system is mimicking such a system with the advantage of providing a handle for tuning a strong field induced split between $|\bar{1}\rangle$ and $|1\rangle$.

**Conclusion**

Utilizing non-perturbative hydrodynamic-Maxwell-Bloch approach we examined properties of the SHG process in a system comprised of two-level emitters strongly coupled to a localized plasmon mode supported by a periodic array of nano-holes in a thin metal film. It is shown that the effective second-order susceptibility exhibits the Rabi splitting at both one- and two-photon peaks. We proposed a simple analytical three-level quantum model, allowing to set on a firm basis the following main findings:

1) The Rabi splitting observed for both one- and two-photon transitions provides the evidence for strong field interaction between plasmonic and molecular parts of the system. Whereas occurrence of such splitting and its sensitivity to the lattice period have been observed and reported for linear properties, it is to our knowledge a first prediction in the context of nonlinear properties. Such features are recognized in both approaches, at comparable spectral positioning and intensity levels.

2) A set of four resonances corresponding to one- and two-photon driven transitions from the ground level to the two plasmon-polariton branches are evidenced in the simulated SHG spectrum for both models.

3) Polarization properties from the simulation properties show well defined threefold and fourfold patterns depending on the experimental consideration. These are accounted for in simple algebraic and geometric terms in the appendix.

Our theoretical studies are surely not to be seen by any means as replacement for actual experiments currently under way but can be considered of a high predictive value in terms of interesting and possibly unexpected phenomena well worthy of a subsequent experimental endeavor. Our results are to be seen as a preliminary step that paves the way onto the design, nanofabrication and optical testing of original configurations of considerable fundamental and applied interests.

**Acknowledgements**

This work was sponsored by the Air Force Office of Scientific Research under Grant No. FA9550-19-1-0009. The authors also acknowledge computational support from the Department of Defense High Performance Computing Modernization Program.

**Appendix A: the fourfold symmetry patterns in SHG angular diagrams**

We proceed to clarify here to account for the patterns of SHG angular polarization plots (APP) from nonlinear entities with threefold symmetry.[35] Such patterns may exhibit four-fold and six-fold, depending on the experimental configuration used, namely the respective orientations of the incoming $\omega$ polarizer versus the $2\omega$ outgoing analyzer, together with the sample rotation scheme, either rotating in-between a fixed polarizer-analyzer configuration or conversely with a fixed sample in-between a rotating polarizer-analyzer configuration. As expected from adequate overall symmetry considerations that encompass both the intrinsic symmetry of the microscopic nonlinear entity at stake together with that of the measurement configuration at the macroscopic laboratory scale, the former microscopic symmetry can be lowered or modified by its convolution with the macroscopic level. Such a response has been already observed in many instances from optically poled polymers[36] to molecular crystals[37] or semiconductor nano-crystals with tetrahedral symmetry,[38] onto metallic holes or metallic nano-particles with triangular shapes.[14,39,40]

We shall further explain here the counter-intuitive difference between the three-fold symmetry shape of the nonlinear entities and the four-fold symmetry of the resulting SHG angular polarization plot in relation with the chosen polarizer-sample-analyzer (PSA) experimental configuration. A different set-up configuration in which the sample is rotating leads to a more intuitive three-fold symmetric APP polar plots that directly reflects the symmetry of the sample (SI).

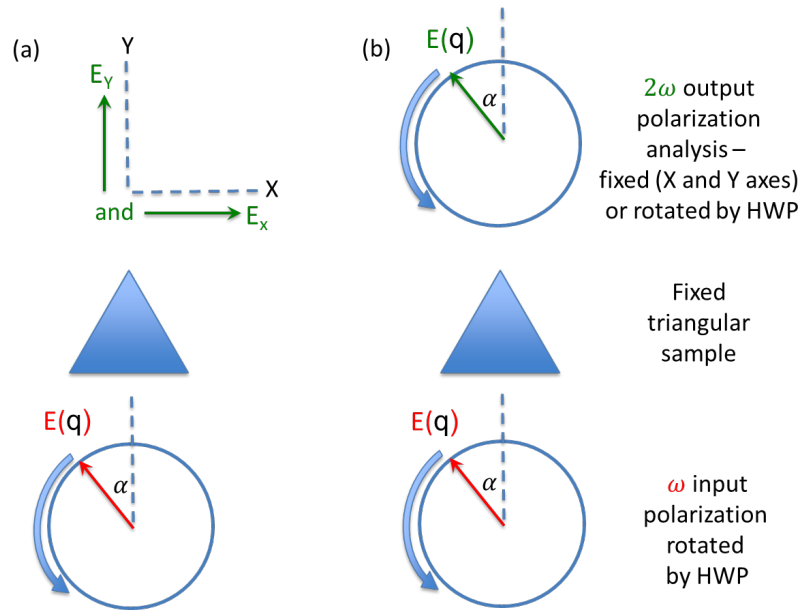

**Fig. A1:** (a) represents the configuration with a rotating fundamental beam polarizer, and fixed harmonic analyzers respectively along *x* and *y*. (b) has been demonstrated elsewhere[41, 42] comprising co-rotating parallel linear polarizer and analyzer. In both configurations, the sample is fixed with its symmetry elements as per the central triangles, regardless of scale considerations.

We shall consider in the following the two configurations depicted in Fig. A1 and provide a tensor-based justification for their APP's as well as a more physically intuitive rationale. Our ultra-thin sample can be considered to a good approximation as quasi two-dimensional thereby cancelling all tensor coefficients with a *z* out-of-plane index, thereby limiting the number of non-vanishing tensor coefficients to those with coefficients limited to combinations of the sole *x* and *y* indices. Under this approximation, only eight non-zero tensor coefficients survive, out of which four are independent coefficients of $\chi^{(2)}_{ijk}$ as a result of index permutation symmetry

$$\chi^{(2)}_{yyy}, \chi^{(2)}_{xxx}, \chi^{(2)}_{xxy} = \chi^{(2)}_{yxx} = \chi^{(2)}_{xyx}, \chi^{(2)}_{xyy} = \chi^{(2)}_{yxy} = \chi^{(2)}_{yyx}. \tag{A1}$$

In the case of isosceles triangular symmetry, an additional mirror plane symmetry perpendicular to the symmetry axis of the triangle leads to the cancellation of additional tensor elements which are anti-symmetric with respect to the mirror plane, e.g. featuring an odd number of *x* or *y* coefficients, namely

$$\chi^{(2)}_{xxx} = \chi^{(2)}_{xyy} = 0. \tag{A2}$$

This further reduces the number of non-zero coefficients from eight to four coefficients, out of which two only are independent

$$\chi^{(2)}_{yyy}, \chi^{(2)}_{yxx} = \chi^{(2)}_{xxy} = \chi^{(2)}_{xyx}. \tag{A3}$$

Further increase of symmetry elements from that of isosceles triangles to equilateral triangles, that is from $D_{2h}$ to $D_{3h}$ imposes an additional relation which results from the very definition of octupolar symmetry, namely, the cancelation of all dipolar tensors attached to the entity of interest. Such symmetry induced cancellation apply to proper physical vectors such as their permanent dipole moment, or adequate combinations of higher order tensor coefficients behaving as vectors under point-wise rotations. In the case of a three rank-order tensor such as the $\chi^{(2)}_{ijk}$ of interest here, the resulting $\vec{\chi}^{(2)}_{i}$ irreducible vector component of the full $\chi^{(2)}_{ijk}$ tensor is given by $\vec{\chi}^{(2)}_{i} = \chi^{(2)}_{ijj}$ (adopting Einstein's convention for repeated indices summation). This leads here to $\chi^{(2)}_{ijj} = 0 = 0$ for all cartesian indices *I = x, y, z*. Implementation of this condition in the case of a planar three-fold object leads to

$$\chi^{(2)}_{yxx} = \chi^{(2)}_{xyx} = \chi^{(2)}_{xxy} = -\chi^{(2)}_{yyy} = \chi^{(2)}_{\parallel}. \tag{A4}$$

This leaves only four mutually dependent non-zero coefficients, which all relate to a single independent $\chi^{(2)}_{\parallel}$ coefficient, the ensuing relations between the transverse incoming field, $\vec{E}$, and the induced polarization components being:

$$P_x = \vec{P}^{2\omega} \cdot \vec{e}_x = \left(\chi^{(2)}_{xyx} + \chi^{(2)}_{xxy}\right) E_x E_y = 2\chi^{(2)}_{\parallel} E_x E_y,$$
$$P_y = \vec{P}^{2\omega} \cdot \vec{e}_y = \chi^{(2)}_{yyy} E_y^2 + \chi^{(2)}_{yxx} E_x^2 = \chi^{(2)}_{\parallel}\left(E_x^2 - E_y^2\right).$$
(A5)

The projection operations along the unit vectors $\vec{e}_x$ and $\vec{e}_y$ unit vectors is the mathematical expression of our experimental configuration (a) in Fig. A1, whereby the output harmonic polarization analyzers are set along those same $\vec{e}_x$ and $\vec{e}_y$ orientations. This is an important remark that will further account for the fourfold symmetry of the measured APP's that convolves the symmetry of the nonlinear entity proper, that is three-fold, together with that for the chosen instrumental PSA configuration.

Taking a linearly polarized input beam with a transverse fundamental field making an angle $\alpha$ with the x-axis, with components $E_x = E_0 \cos\alpha$ and where $E_0$ stands for the field modulus. The second order harmonic polarization is then given by

$$P_x^{2\omega}(\alpha) = \chi^{(2)}_{\parallel} E_0^2 \sin(2\alpha),$$
$$P_y^{2\omega}(\alpha) = \chi^{(2)}_{\parallel} E_0^2 \cos(2\alpha).$$
(A6)

The far-field SHG intensity is, to a multiplicative constant, the above squared expressions of the polarization components

$$I_x^{2\omega}(\alpha) \propto \left(\chi^{(2)}_{\parallel}\right)^2 \left(I^\omega\right)^2 \sin^2(2\alpha),$$
$$I_y^{2\omega}(\alpha) \propto \left(\chi^{(2)}_{\parallel}\right)^2 \left(I^\omega\right)^2 \cos^2(2\alpha).$$
(A7)

These harmonic intensities reported here in Fig. A1 correspond to experimental configurations with a rotating input linear polarizer on a rotated mount allowing to cover a full circular rotation in the $0 \leq \alpha \leq 2\pi$ interval, and two fixed linear harmonic analyzers along x and y, the sample itself being set on a fixed mount. These expressions are in agreement with the experimentally observed as well as calculated fourfold symmetry pattern in Figure 2 (c and d) as from their respective $\sin^2(2\alpha)$ and $\cos^2(2\alpha)$ angular dependences. It also reflects the observed $\pi/4$ angular shift in-between the two x- and y-analyzed APP's, namely

$$I_x^{2\omega}(\alpha) = I_y^{2\omega}\left(\frac{\pi}{4} - \alpha\right)$$
(A8)

is a simple consequence. Figure A2 shows specific configurations that illustrate the ensuing four-fold pattern of the APP for this configuration.

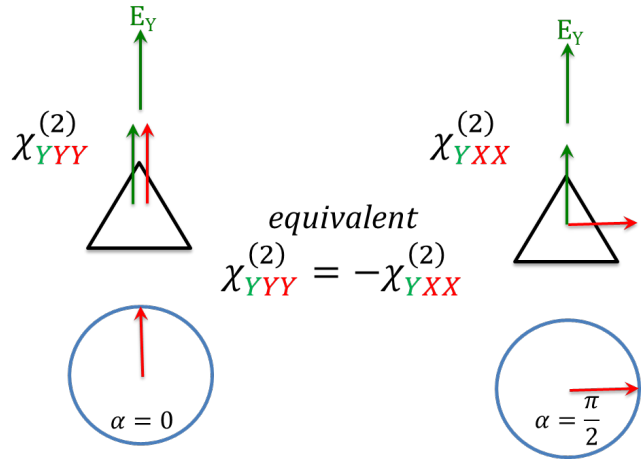

**Fig. A2**. Two specific configurations in (a) (from Fig. A1) with parallel and crossed polarizer and analyzer, the latter set along Y in both cases. These can be viewed as equivalent due to the identity of the mediating $\chi^{(2)}_{YXX} = -\chi^{(2)}_{YYY} = \chi^{(2)}_{\parallel}$ tensor coefficients in both cases (to a minus sign that disappears after squaring).

One can further illustrate the $\pi/4$ shift when the analyzer is set along $x$, with respect to the former configuration, as shown in Fig. A3.

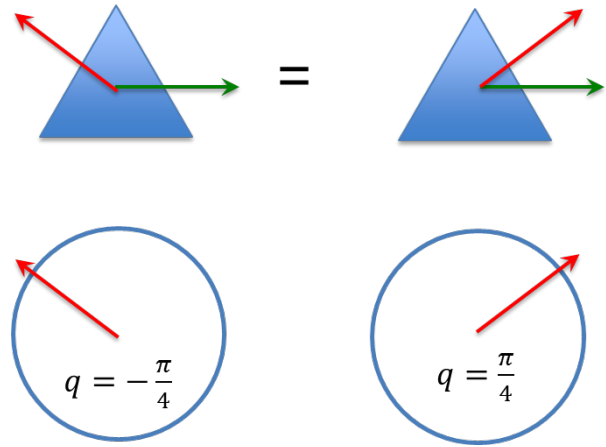

**Fig. A3**. Two specific configurations in (a) (from Fig. A1) with parallel and crossed polarizer and analyzer, the latter set along Y in both cases. These can be viewed as equivalent due to the identity of the mediating $\chi^{(2)}_{yxx} = -\chi^{(2)}_{yyy} = \chi^{(2)}_{\parallel}$ tensor coefficients in both cases (to a minus sign that disappears after squaring).

For the sake of illustrating the contribution of the PSA configuration onto the symmetry pattern of the APP in that configuration let us now investigate an alternative one where the linearly polarized input polarizer and output analyzer are parallel and co-rotating in the $0 \leq \alpha \leq 2\pi$ interval as shown on Fig. A4 whereby the $\alpha = 0, \frac{2\pi}{3}, \frac{4\pi}{3}$ angular positions are being singled-out. It comes clearly out from inspection of Fig. A4 that those three positions (as well as any other three related to this one by an arbitrary in-plane rotation) are equivalent and thus suggesting that the QAP must exhibit a threefold symmetry pattern, or rather a sixfold one when squaring the outgoing field into the corresponding considering intensity. Such a configuration has been used for example in Refs. [41, 42]

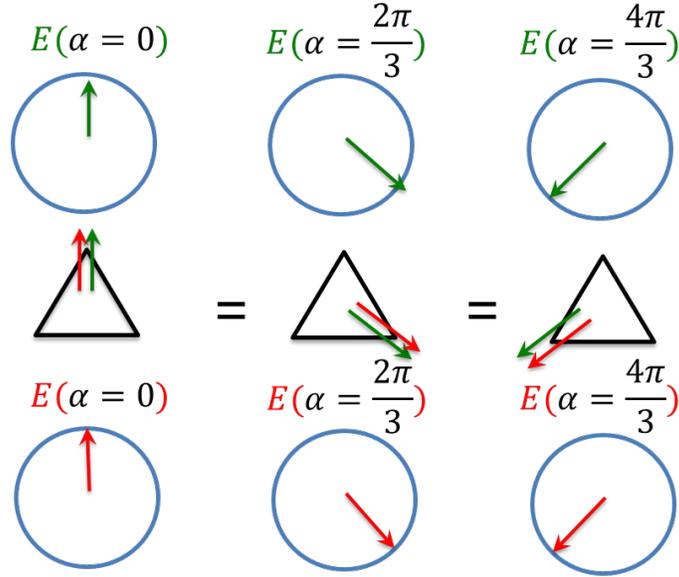

**Fig. A4**. co-rotating parallel polarizer and analyzer at three symmetric angular positions in configuration (b) from Fig. A1.

The above projections along the $\vec{e}_x$ and $\vec{e}_y$ unit vectors become then irrelevant and we turn now to the more appropriate projection onto the rotating unit vector $\vec{u} = \vec{e}_x \cos\alpha + \vec{e}_y \sin\alpha$

$$P_u(\alpha) = \vec{P}^{2\omega} \cdot \vec{u}(\alpha) = \chi_{\|}^{(2)} E_0^2 \sin(3\alpha),$$
$$I_u^{2\omega}(\alpha) \propto \left(\chi_{\|}^{(2)} I^{\omega}\right)^2 \sin^2(3\alpha). \tag{A9}$$

This PSA configuration leads to the more "natural" six-old (due to the squaring from the threefold $P_u(\alpha)$ to a then sixfold $I_u^{2\omega}(\alpha)$ angular dependence).

Likewise, one can straightforwardly show that the alternative co-rotating *perpendicular* polarizer-analyzer configuration (Fig. 2b) leads to a sixfold symmetry pattern rotated by 30° with respect to the former parallel *perpendicular* polarizer-analyzer configuration (Fig. 2a).